\documentclass[debug]{rmaa}

\usepackage{longtable,lscape}
\usepackage{multirow}
\usepackage{natbib}

\title{Planetary nebulae in the inner Milky Way II: the Bulge-Disk transition}
\author{
  O. Cavichia,\altaffilmark{1} 
  R. D. D. Costa,\altaffilmark{1} 
  and W. J. Maciel\altaffilmark{1}}

\altaffiltext{1}{Instituto de Astronomia, Geof\'\i sica e Ci\^encias
                 Atmosf\'ericas, Universidade de S\~ao Paulo, Brazil.}

\shortauthor{Cavichia, Costa \& Maciel}
\shorttitle{Planetary nebulae in the inner Milky Way II}

\fulladdresses{
\item Departamento de Astronomia, Instituto de Astronomia, Geof\'isica e Ci\^encias Atmosf\'ericas, Universidade de S\~ao Paulo, Rua do Mat\~ao, 1226, Cidade Universit\'aria, 05508-900 S\~ao Paulo-SP, Brazil.
}

\listofauthors{O. Cavichia, R. D. D. Costa, \& W. J. Maciel}
\indexauthor{Cavichia, O.}
\indexauthor{Costa, R. D. D.}
\indexauthor{Maciel, W. J.}

\abstract{In this work, a sample of planetary nebulae located in the inner-disk and bulge of the Galaxy is used in order to find the galactocentric distance which better separates these two populations, from the point of view of abundances. Statistical distance scales were used to study the distribution of abundances across the disk-bulge interface. A Kolmogorov-Smirnov test was used to find the distance in which the chemical properties of these regions better separate.

The results of the statistical analysis indicate that, on the average, the inner population has lower abundances than the outer.  Additionally, for the $\alpha$-elements abundances, the inner population does not follow the disk radial  gradient towards the galactic center. Based on  our results, we suggest a bulge-disk interface at 1.5 kpc, marking the transition between the bulge and inner-disk of the Galaxy, as defined by the intermediate mass population.}

\resumen{}

\addkeyword{Galaxy: abundances}
\addkeyword{Galaxy: evolution}
\addkeyword{planetary nebulae: general}
\addkeyword{techniques: spectroscopic}

\begin{document}
\maketitle

\section{Introduction}
\label{sec:intro}

Chemical abundances of Planetary Nebulae (PNe) are an important tool to investigate the chemical evolution of the Galaxy. As a final stage of evolution of low and intermediate-mass stars ($\sim1-8\mbox{ M}_\odot$), the abundances of  $\alpha$-elements found in PNe usually are not modified by the evolution of the progenitor star. In this sense, chemical abundances of PNe can be used to study the chemical evolution of galaxies as done by \citet{maciel05,maciel06,fu09, stangh10}. 

The chemical abundances obtained from PNe are relatively accurate and the nebulae can be observed at large distances in the Galaxy, even in regions obscured by the dust in the galactic disk, since they have bright emission lines, such as  [\ion{O}{III}]$\lambda500.7\mbox{nm}$ and $\mbox{H}\alpha$. Therefore PNe are useful to study the pattern of abundances in the inner-disk and bulge of the Galaxy, where the extinction in the optical range is very severe \citep[see for example][]{escudero01,escudero04,chiap09,cavichia10}. On the other hand, PNe distances are still subject of discussion. Only in a few cases the distances can be determined by a direct method, such as for nearby PNe, which have the distances determined by trigonometric parallax, or in cases where there is a main sequence binary companion. In the cases were accurate individual distances cannot be determined, alternative methods were developed in order to obtain reliable distances. In these cases, they have their distances derived from nebular properties (see e.g. \citealp{maciel80,cahn92,stangh08}). These methods are called statistical methods and, in this case, the distances obtained are called statistical distances. In spite of the uncertainties, statistical distance scales are still the best tool to study the chemical abundance patterns in the Galaxy from the point of view of PNe, as done by \citet{maciel99,maciel06,perinotto06} and recently by \citet{guten08};\\ \citet{stangh10,henry10}.

From the point of view of the chemical evolution of the Galaxy, the bulge and the disk display different chemical abundance patterns such as the radial abundance gradients in the disk \citep{carigi05,daflon04,andri04,maciel05,maciel06}, or the wide abundance distribution in the bulge \citep{rich88,zoccali03,zoccali06}. Also, the bulge and the disk may have different evolution histories, described for example by the multiple infall scenario \citep{costa05,costa08} or by the inside out formation model \citep{chiap01}, respectively. As a consequence, we expect that these differences are reflected on the chemical properties of each component. 

Additionally, the gas density, and consequently the star formation rate (SFR), also decrease in the inner Galaxy as found by \citet{portinari99}. They attribute this result to the influence of the galactic bar in the first 3 kpc of the Galaxy (see for example \citealp{gerhard02,binney09}). The decreasing of the SFR is reflected on the $\alpha$-element abundances in PNe, as a lower SFR limits the total mass ejected by the progenitor star, and consequently the chemical enrichment of the ISM decreases throughout the succeeding stellar generations. The galactic evolution models also point out to a severe decrease in the SFR and chemical abundances at radii smaller than 3 kpc \citep{porti00}. Other chemical evolution models also predict a flatter abundance gradient in the inner-disk such as those by \citet{hou00} and \citet{molla05}.

Some recent works seem to support this scenario, based on the fact that towards the Galactic center, stars do not follow the chemical abundance gradient found in the disk \citep{cunha07}. \citet{davies09} studying the chemical abundance patterns of the Scutum red supergiant clusters, located at the end of the Galactic bar, find that there is a minimum of abundances in this region, compared to other azimuthal angles.

From the point of view of PNe, the results of \citet{guten08} also point to a discontinuity in the abundance gradient towards the galactic center, in the sense that the abundances of bulge PNe do not follow the trend of those from the disk, as can be seen in their figure 4. \citet{gorny04} also notice that the oxygen gradient of the PNe population in the galactic disk flattens in the most internal parts of the Galaxy, and may even change sign. On the other hand, \citet{chiap09} made a comparison between abundances from PNe located at the bulge, inner-disk and Large Magellanic Cloud. Their results do not show any clear difference between bulge and inner-disk objects. 

Other previous studies of the Galactic bulge based on abundances of PNe such as \citet{ratag92,cuisin00,escudero01,escudero04,exter04}, find that bulge PNe have an abundance distribution similar to disk PNe, showing that He, O, Si, Ar, and Ca have a normal abundance pattern, favouring therefore a slower Galactic evolution than that indicated by stars. In conclusion, the distribution of chemical abundances in the inner region of the Galaxy is still an open question, especially regarding the bulge-disk connection. On the other hand, a detailed comparison of the bulge and inner-disk populations is important in order to characterize the galactic bulge as a ``pseudobulge'', as suggested by photometric and kinematic evidences \citep{howard09, shen10}, in opposition to a classical bulge as suggested by \citet{binney09}.

Both from modelling and observational results, we expect that exist a galactocentric distance which better separates the populations of the bulge and inner-disk from the point of view of chemical abundances. However, such transition is still to be determined from the point of view of chemical abundances. The goal of this paper is to perform a statistical analysis using the new data provided by \citet{cavichia10}, hereafter Paper I,  and additional data from the literature, in order to characterize the interface between the bulge and inner-disk of the Galaxy, as defined by the PNe population. Taking into account both the abundance distributions and the individual distances for a sample of PNe,  we are able to characterize the interface by establishing at which galactocentric distance bulge and disk characteristics better separate.

This paper is organized as follows: in \S~\ref{sec:sample_distances} the details of the
sample and the new distances to our objects are presented. In \S~\ref{sec:method} the method used in order to characterize the bulge-disk transition is described. In \S~\ref{sec:discussion} we discuss the results and compare them with other results in the literature. Finally, in \S~\ref{sec:summary} the main conclusions are presented.

\section{The Sample and new distances}
\label{sec:sample_distances}

\subsection{The Sample}

An extensive number of statistical scales have been proposed in the literature, and two of them have been adopted in order to study the chemical abundance distribution in the inner Galaxy: \citet{stangh08}, hereafter SSV08, and \citet{zhang95}, hereafter Z95, distance scales. These scales are based on the Shklovsky method but with the introduction of a relationship between the ionized mass and the radius of the nebula, as proposed by \citet{maciel80}. SSV08 is a recent revision of the \citet{cahn92}, hereafter CKS92, scale based on the physical properties of the PNe in the Magellanic Clouds, whose distances are accurately known. As the authors argue, this new distance scale is the most reliable to date.

The selection criteria applied to search for objects in the two catalogues (SSV08 and Z95) were the following: we selected those PNe which had 5 GHz fluxes lower than 100 mJy, optical diameters lower than 12 arcsec, and galactic coordinates within the range $|\ell| \leq 10^{\circ}$ and $|b| \leq 10^{\circ}$. The galactic coordinates were used to take into account only the PNe which are in the galactic center direction. The combination of the other two criteria leads to the rejection of about 90--95\% of the PNe which are in the galactic center direction, but have heliocentric distances lower than 4 kpc (cf. \citealt{stasinska98}). These criteria are commonly used by other authors to select bulge PNe (e.g. \citealt{exter04} and \citealt{chiap09}), so that most of the objects considered in this work should be at or near the bulge. 

In order to study the chemical abundance distribution in the inner Galaxy, we have made use of the data published by our group \citep{cavichia10,escudero04,escudero01}. We also searched for chemical abundances of PNe located in the bulge and inner-disk of the Galaxy in the following works: \citet{ratag97,exter04,gorny04,perinotto04,pacheco92,koppen91,samland92,cuisin00}. All these works have the same region of interest, and, besides that, they have significant and homogeneous samples, which is very important in multi-object study, as we propose in this work. For more details on these works see Paper I. Including these additional objects to our original sample in Paper I, our database amounts to 140 objects, as given in table \ref{tab:all_dist}. In this table, the columns give the PN G number, the name of the nebula, the galactocentric distance in kpc for the SSV08 and the Z95 distance scales, respectively. An asterisk marks those PNe whose distances were derived in the present work. The galactocentric distances of the objects located beyond the galactic bulge are given with a negative sign in table \ref{tab:all_dist}. The extinction in this region is very high, and the uncertainties in the distances are large,  reaching 30\% in some cases. For example, an object with a heliocentric distance of 8 kpc in this direction for both scales will have a true heliocentric distance between 5.6 and 10.4. Therefore, it is unlikely that these objects are really located far beyond the galactic bulge, so that we assume that they actually belong to the bulge population. The implications of this assumption will be further discussed in section 3.3. More details on the new distances and the determination of the galactocentric distances are given in section \ref{sec:new_dist}.

\begin{center}
\footnotesize
\begin{longtable}{lccc}
\caption{\label{tab:all_dist}Galactocentric distances of PNe in the sample.}\\
\toprule
PN G  & Name & $\mbox{R}_{\mbox{\tiny SSV08}}$ (kpc)  & $\mbox{R}_{\mbox{\tiny Z95}}$ (kpc)\\ 
\midrule
\endfirsthead
\caption{continued.}\\
\toprule
PN G  & Name & $\mbox{R}_{\mbox{\tiny SSV08}}$ (kpc)  & $\mbox{R}_{\mbox{\tiny Z95}}$ (kpc)\\ 
\midrule
\endhead
\bottomrule
\endfoot

000.1-01.1	&	M3-43   	&	--	&	1.81	\\
000.2-01.9	&	M2-19   	&	-0.30	&	0.63	\\
000.3-04.6	&	M2-28   	&	--	&	-0.70	\\
000.4-02.9	&	M3-19	&	0.31	&	-0.73	\\
000.7-02.7	&	M2-21	&	-2.20	&	--	\\
000.7-03.7	&	M3-22	&	0.50	&	-0.63	\\
000.7+03.2	&	Hen 2-250 	&	--	&	1.07	\\
000.9-04.8	&	M3-23	&	3.78	&	3.45	\\
001.0+01.9 	&	K1-4	&	6.06	&	--	\\
001.2-03.0	&	H1-47	&	--	&	-2.07	\\
001.2+02.1	&	Hen 2-262 	&	-1.57	&	1.39	\\
001.4+05.3	&	H1-15	&	-1.29	&	0.73	\\
001.7-04.6	&	H1-56   	&	--	&	-3.06	\\
001.7+05.7	&	H1-14	&	1.73	&	2.24	\\
002.0-02.0*	&	H1-45	&	-7.12	&	-0.36	\\
002.0-06.2	&	M2-33   	&	-1.09	&	-1.02	\\
002.1-02.2	&	M3-20	&	0.44	&	2.71	\\
002.1-04.2	&	H1-54   	&	-4.39	&	-0.57	\\
002.2-02.7	&	M2-23	&	3.75	&	-0.51	\\
002.2-09.4	&	Cn1-5	&	2.83	&	3.70	\\
002.4+05.8 	&	NGC6369	&	6.92	&	--	\\
002.6-03.4	&	M1-37   	&	--	&	-0.62	\\
002.6+08.1	&	H1-11	&	1.19	&	1.30	\\
002.7-04.8	&	M1-42   	&	2.53	&	3.15	\\
002.7+01.6	&	H2-20   	&	--	&	0.53	\\
002.8-02.2	&	Pe2-12	&	--	&	-4.12	\\
003.1+02.9	&	Hb4	&	2.98	&	--	\\
003.1+03.4	&	H2-17   	&	--	&	-0.98	\\
003.2-06.1	&	M2-36	&	2.03	&	1.29	\\
003.3-04.6 	&	Ap1-12	&	3.42	&	--	\\	
003.4-04.8	&	H2-43   	&	--	&	3.03	\\
003.5-04.6	&	NGC6565 	&	3.37	&	3.68	\\
003.6-02.3	&	M2-26	&	-0.40	&	-1.46	\\
003.7-04.6	&	M2-30	&	-1.55	&	-0.68	\\
003.8-04.3	&	H1-59   	&	-1.53	&	-3.18	\\
003.8-04.5	&	H2-41	&	-0.85	&	-6.23	\\
003.8+05.3	&	H2-15	&	-7.40	&	-8.96	\\
003.9-02.3	&	M1-35   	&	1.93	&	3.38	\\
004.0-03.0	&	M2-29	&	-0.75	&	-2.01	\\
004.0-05.8	&	Pe1-12	&	1.66	&	-2.33	\\
004.1-04.3	&	H1-60   	&	-2.24	&	-0.82	\\
004.6+06.0	&	H1-24   	&	--	&	0.89	\\
004.8-05.0	&	M3-26   	&	0.82	&	-0.82	\\
004.8+02.0	&	H2-25	&	-3.45	&	-4.22	\\
004.9+04.9	&	M1-25	&	1.50	&	3.11	\\
005.0+04.4	&	H1-27   	&	0.65	&	-2.94	\\
005.2+05.6	&	M3-12   	&	--	&	1.24	\\
005.5-02.5* 	&	M3-24	&	1.34	&	--	\\
005.7-05.3	&	M2-38	&	1.24	&	-1.01	\\
005.8-06.1	&	NGC6620	&	-1.19	&	1.61	\\
005.8+05.1 	&	H2-16	&	2.81	&	--	\\
006.0-03.6	&	M2-31   	&	--	&	3.14	\\
006.1+08.3*	&	M1-20	&	-1.12	&	2.34	\\
006.4-04.6*	&	Pe2-13	&	1.92	&	--	\\
006.4+02.0	&	M1-31   	&	2.94	&	4.10	\\
006.8-03.4*	&	H2-45	&	1.89 & 	--      \\
006.8+04.1	&	M3-15   	&	--	&	3.94	\\
007.0-06.8	&	VY2-1   	&	0.86	&	2.57	\\
007.0+06.3*	&	M1-24	&	-1.00	&	--	\\
007.1-06.0	&	H1-66   	&	--	&	-1.70	\\
007.2+01.8* 	&	Hb6	         & 	3.78	&	--	\\
007.5+07.4	&	M1-22   	&	--	&	-4.02	\\
007.8-04.4	&	H1-65   	&	--	&	-2.25	\\
008.0+03.9 	&	NGC6445	&	6.64	&	--	\\
008.1-04.7	&	M2-39   	&	-4.02	&	-2.21	\\
008.2-04.8	&	M2-42   	&	-1.48	&	0.66	\\
008.2+06.8*	&	Hen 2-260	&	-12.23	&	-3.62	\\
008.3-07.3*	&	NGC6644 	&	2.18	&	4.01	\\
009.0+04.1	&	Th4-5   	&	--	&	1.69	\\
009.4-05.0	&	NGC6629	&	5.67	&	--	\\
009.4-09.8	&	M3-32	&	1.85	&	1.57	\\
009.6-10.6	&	M3-33	&	-1.45	&	--	\\
009.8-04.6	&	H1-67	&	0.64	&	0.80	\\
010.4+04.5	&	M2-17   	&	--	&	0.85	\\
010.7-06.4	&	IC4732	&	1.51	&	3.22	\\
010.7-06.7 	&	Pe1-13	&	-1.37	&	--	\\
010.7+07.4* 	&	Sa2-230	&	1.41	&	--	\\
010.8-01.8	&	NGC6578	&	4.42	&	--	\\
011.0+05.8	&	NGC6439	&	1.89	&	--	\\
350.5-05.0*	&	H1-28       &	-1.93 &	--	\\
350.9+04.4	&	H2-1	&	2.45	&	--	\\
351.1+04.8	&	M1-19	&	-2.75	&	0.71	\\
351.2+05.2	&	M2-5    	&	--	&	0.80	\\
351.6-06.2*	&	H1-37	&	-1.81	&	--	\\
352.0-04.6	&	H1-30   	&	--	&	2.88	\\
352.1+05.1	&	M2-8	&	-1.39	&	1.05	\\
352.6+03.0	&	H1-8	&	-0.91	&	--	\\
353.7+06.3	&	M2-7	&	1.19	&	-1.00	\\
354.2+04.3	&	M2-10   	&	--	&	-2.73	\\
355.1-02.9	&	H1-31   	&	--	&	-3.78	\\
355.1-06.9	&	M3-21   	&	--	&	2.55	\\
355.4-02.4	&	M3-14	&	2.33	&	1.58	\\
355.4-04.0 	&	Hf2-1	&	1.92	&	--	\\
355.7-03.0	&	H1-33	&	-4.05	&	-1.23\\
355.7-03.4*	&	H1-35   	&	2.15	&	2.03	\\
355.9-04.2	&	M1-30   	&	--	&	2.18	\\
355.9+03.6*	&	H1-9	&	2.19	&	--	\\
356.2-04.4	&	Cn2-1	&	-0.56	&	1.91	\\
356.3-06.2 *	&	M3-49	&	-5.33	&	--	\\
356.5-02.3	&	M1-27   	&	3.41	&	4.27	\\
356.5-03.9	&	H1-39	&	-7.28	&	-2.08	\\
356.7-04.8	&	H1-41	&	2.37	&	1.51	\\
356.8-05.4* 	&	H2-35	&	1.86	&	--	\\
356.9-05.8	&	M2-24	&	-1.52	&	-3.86	\\
356.9+04.4	&	M3-38   	&	-6.07	&	-0.65	\\
356.9+04.5	&	M2-11   	&	--	&	1.08	\\
357.1-04.7	&	H1-43	&	-5.33	&	-3.59	\\
357.1+03.6	&	M3-7	&	1.73	&	2.62	\\
357.2-04.5	&	H1-42	&	1.94	&	3.22	\\
357.2+07.4	&	M4-3    	&	--	&	-0.37	\\
357.3+03.3	&	M3-41   	&	--	&	3.87	\\
357.3+04.0	&	H2-7    	&	--	&	-1.32	\\
357.4-03.2	&	M2-16	&	1.05	&	2.18	\\
357.4-03.5	&	M2-18	&	-7.74	&	-1.48	\\
357.4-04.6	&	M2-22	&	-0.79	&	-1.97	\\
357.5+03.2	&	M3-42	&	1.17	&	-1.78	\\
357.6+01.7	&	H1-23	&	-1.74	&	1.89	\\
357.6+02.6	&	H1-18   	&	--	&	0.72	\\
357.9-05.1	&	M1-34   	&	--	&	2.15	\\
358.2+03.5	&	H2-10	&	-5.35	&	-0.50	\\
358.2+03.6	&	M3-10   	&	-1.40	&	1.72	\\
358.2+04.2	&	M3-8	&	0.96	&	0.78	\\
358.3-02.5	&	M4-7    	&	1.66	&	2.81	\\
358.3+03.0*	&	H1-17	&	0.42	&	-0.59	\\
358.5-04.2*	&	H1-46	&	-0.75	&	1.04	\\
358.6+01.8	&	M4-6    	&	--	&	1.76	\\
358.6+07.8	&	M3-36   	&	--	&	-6.05	\\
358.7-05.2*	&	H1-50   	&	-3.03	&	0.79	\\
358.8+03.0 	&	Th3-26	&	0.66	&	--	\\
358.9+03.2	&	H1-20	&	0.64	&	2.91	\\
358.9+03.3*	&	H1-19   	&	-2.95	&	-0.51	\\
359.1-01.7	&	M1-29   	&	--	&	4.70	\\
359.1-02.3	&	M3-16	&	1.30	&	1.50	\\
359.3-03.1	&	M3-17   	&	-3.60	&	-1.10	\\
359.4-03.4	&	H2-33	&	0.54	&	-1.78	\\
359.7-02.6	&	H1-40	&	0.84	&	0.36	\\
359.8+03.7 	&	Th3-25	&	-6.52	&	--	\\
359.8+06.9	&	M3-37	&	-8.69	&	-5.50	\\
359.9-04.5	&	M2-27	&	-4.45	&	2.68	\\
359.9+05.1 	&	M3-9	&	4.70	&	--	\\			     
\end{longtable}
\end{center}

\subsection{New distances to the Planetary Nebulae \label{sec:new_dist}}

Several of the objects for which the chemical abundances were determined in Paper I, do not have distances published by SSV08. In order to include these PNe in our analysis, their distances were estimated from the equations (8a) and (8b) of SSV08. In the cases where the 5 GHz flux is not available, their equivalent 5 GHz flux from H$\beta$ flux was derived using equation (6) in CKS92. The optical thickness parameter was derived from equation (2) in SSV08. Angular radii were obtained in the literature, most of them from \citet{acker92}, and some from more recent works such as \citet{tylenda03} and \citet{ruffle04}. The objects with new distances are marked with an asterisk in table \ref{tab:all_dist}, and the data used in the determination are shown in table \ref{tab:new_dist}. In this table, column 1 gives the PN G number, column 2 the name, column 3 the angular diameter, column 4 the 5 GHz flux, column 5 the optical thickness parameter, and column 6 results the new distances derived for these objects.

\begin{table}[ht!]
  \begin{center}
  \footnotesize
  \caption{New distances.}
  \label{tab:new_dist}
  \begin{tabular}{l c c c c c }
  \toprule
PN G        	&	Name	&	$\theta$ (arcsec)	&	 F (mJy)	&	 $\tau$	&	d (kpc)	\\
 \midrule										
002.0-02.0	&	H1-45	&	0.8	&	20	&	2	&	15.1	\\
005.5-02.5	&	M3-24	&	5.1	&	4.9	&	4.3	&	6.7	\\
006.1+08.3	&	M1-20	&	0.8	&	50	&	1.7	&	8.1	\\
006.4-04.6	&	Pe2-13	&	6.6	&	3.4	&	4.7	&	6.2	\\
006.8-03.4	&	H2-45	&	4.6	&	9.7	&	3.9	&	6.2	\\
007.0+06.3	&	M1-24	&	3.2	&	6.2	&	3.8	&	8.5	\\
007.2+01.8	&	Hb6	&	2.5	&	240	&	2	&	4.3	\\
008.2+06.8	&	Hen 2-260	&	0.5	&	10	&	1.9	&	20.4	\\
008.3-07.3	&	NGC6644	&	1.3	&	100	&	1.8	&	6.1	\\
010.7+07.4	&	Sa2-230	&	5	&	4	&	4.4	&	7.1	\\
350.5-05.0	&	H1-28 	&	3.9	&	1.5	&	4.6	&	9.9	\\
351.6-06.2	&	H1-37	&	4.3	&	1.3	&	4.7	&	9.7	\\
355.7-03.4	&	H1-35 	&	1	&	90	&	1.7	&	5.9	\\
355.9+03.6	&	H1-9	&	3.5	&	30	&	3.2	&	5.9	\\
356.3-06.2	&	M3-49	&	4.9	&	0.2	&	5.7	&	13.2	\\
356.8-05.4	&	H2-35	&	5.4	&	0.9	&	5.1	&	9.1	\\
358.3+03.0	&	H1-17	&	0.5	&	40	&	1.4	&	7.9	\\
358.5-04.2	&	H1-46	&	0.6	&	40	&	1.5	&	8.5	\\
358.7-05.2	&	H1-50	&	0.7	&	30	&	1.8	&	10.9	\\ 
358.9+03.3	&	H1-19	&	0.7	&	30	&	1.9	&	10.9	\\
  \bottomrule					     
  \end{tabular}
 \end{center}
\end{table}

\begin{figure}[ht!]
	\centering
	\includegraphics[width=11cm]{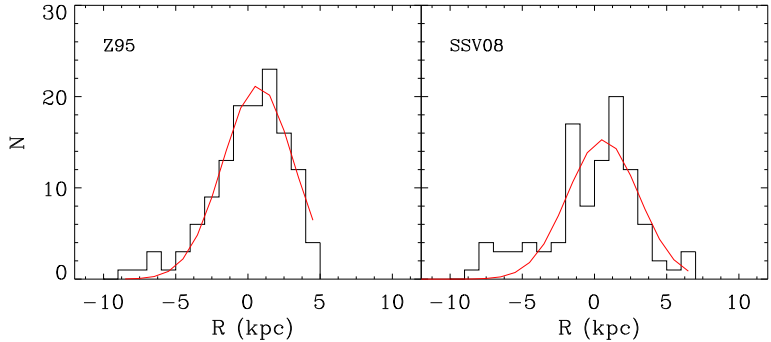}
	\caption{Galactocentric distance distribution for the Z95 distance scale (left) and for the SSV08  distance scale (right). For each histogram a gaussian fit is shown (see the text for more details).}
	\label{hist_dist}
\end{figure}

\subsection{Galactocentric distances and distribution}

Heliocentric distances were converted into galactocentric distances using equation \ref{dist_gal}, where $d$ is the heliocentric distance, $b$ and $\ell$ the Galactic latitude and longitude, respectively, and $R_0$ the solar distance to the Galactic center. In this work we adopted $R_0 = 8 \mbox{ kpc}$, as suggested by recent studies \citep{gillessen09,nishi06}.

\begin{equation}
R = \sqrt{R_0^2 + (d \ cosb)^2 - 2\ R_0\ d\ cosb\ cos\ell},
\label{dist_gal}
\end{equation}

In order to check the distribution of distances for the PNe in our database, figure \ref{hist_dist} shows the distributions of the PNe with respect to the galactocentric distance for both distance scales used in this work, Z95 (left) and SSV08 (right), including the new distances derived in this work. The figure also shows a gaussian fit to each histogram. The mean and standard deviation are ($0.6 \pm 0.3$) kpc and ($2.6 \pm 0.3$) kpc for SSV08 scale, and ($0.6 \pm 0.2$) kpc and ($2.5 \pm 0.2$) kpc for Z95 scale. For both scales, the mean is consistent with the most recent determinations of the solar distance to the Galactic center, which are near 8 kpc, while the widths are larger than the standard deviation expected of 1.2 kpc for the bulge. As discussed by Z95, this dispersion is a convolution of the probably real gaussian distribution of the bulge, with the extra spread introduced by the mass-radius relationship.

\section{Determination of the bulge-disk interface}
\label{sec:method}

\subsection{ Abundance differences between bulge and inner-disk PNe}

In order to study the chemical abundance distribution in the inner Galaxy and determine the bulge-disk interface, we made use of a method similar to that suggested by \citet{maciel06}. First, a galactocentric distance ($\mbox{R}_{\mbox{L}}$) is chosen, which defines a limit for the sample. Then the sample is divided into two groups: group I is composed by PNe with galactocentric distances smaller than the limit previously settled, and group II, composed by PNe with galactocentric distances larger than this limit. For each group, average abundances and the standard error of the mean are calculated for the available elements, namely He, O, N, S, and Ne. Then, the limiting radius $\mbox{R}_{\mbox{L}}$ is varied, considering distances in the interval from 0.1 to 3.6 kpc, in 0.7 kpc steps for each distance scale.

\begin{figure*}[ht!]
	\centering
	\includegraphics[width=6.5cm]{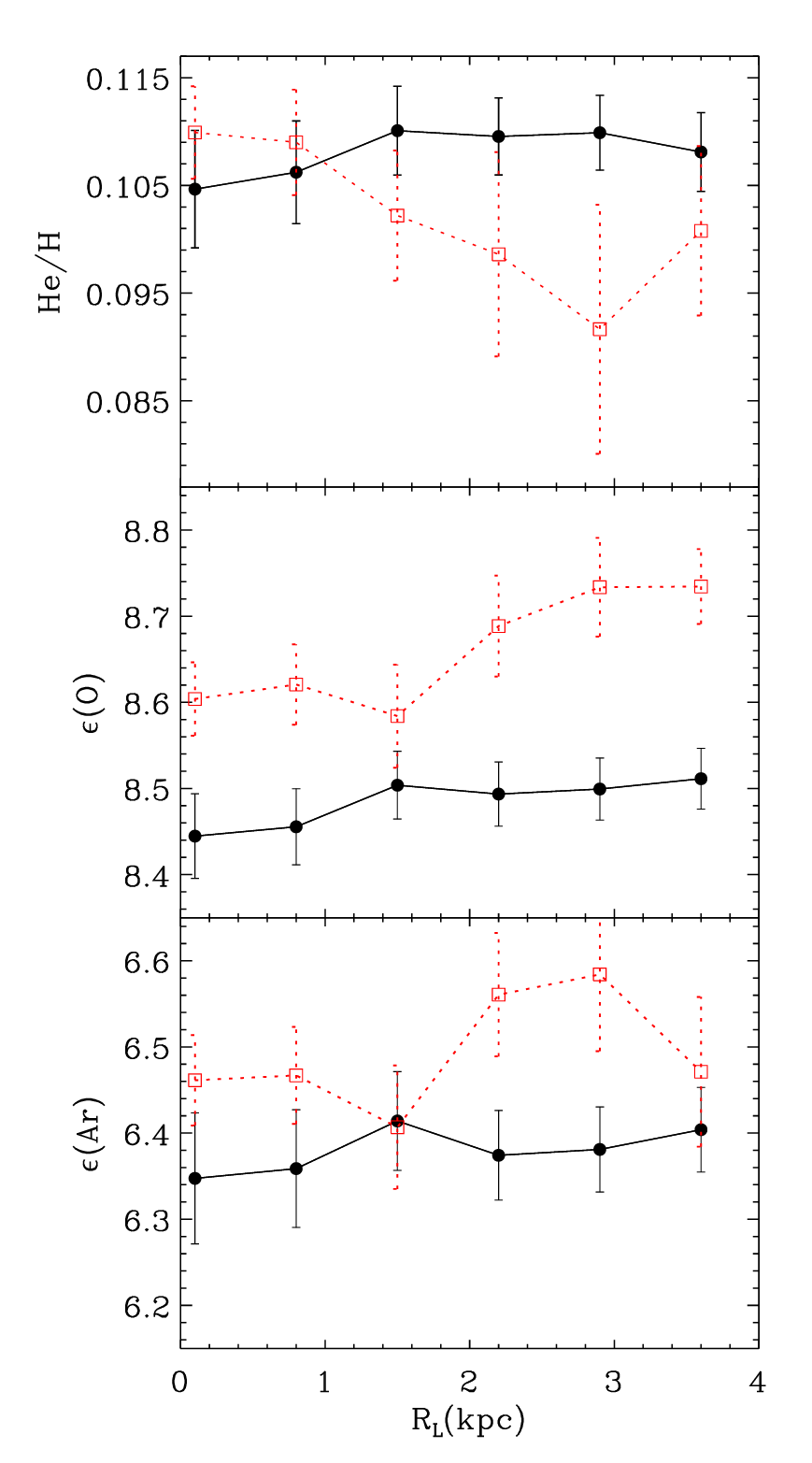}
	\includegraphics[width=6.5cm]{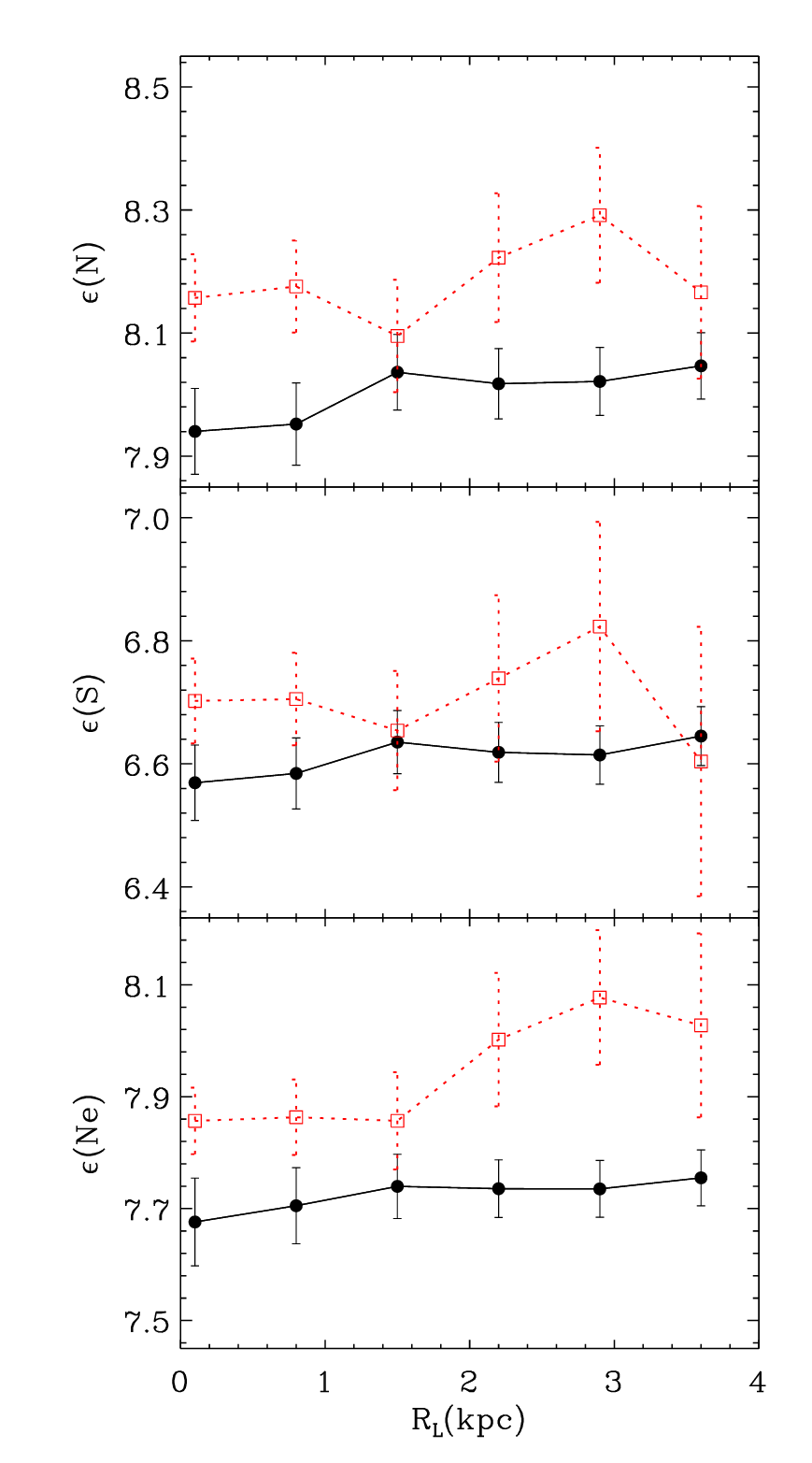}
	\caption{Average abundances of PNe of group I (circles joined by continuous lines) and group II (squares joined by dashed lines) for the SSV08 distance scale. The abscissa gives the limiting galactocentric distance $\mbox{R}_{\mbox{L}}$, considered to be in the range $0-4 \mbox{ kpc}$. Error bars indicate the standard error of the mean for each subsample.}
	\label{stangh}
\end{figure*}

\begin{figure*}[ht!]
	\centering
	\includegraphics[width=6.5cm]{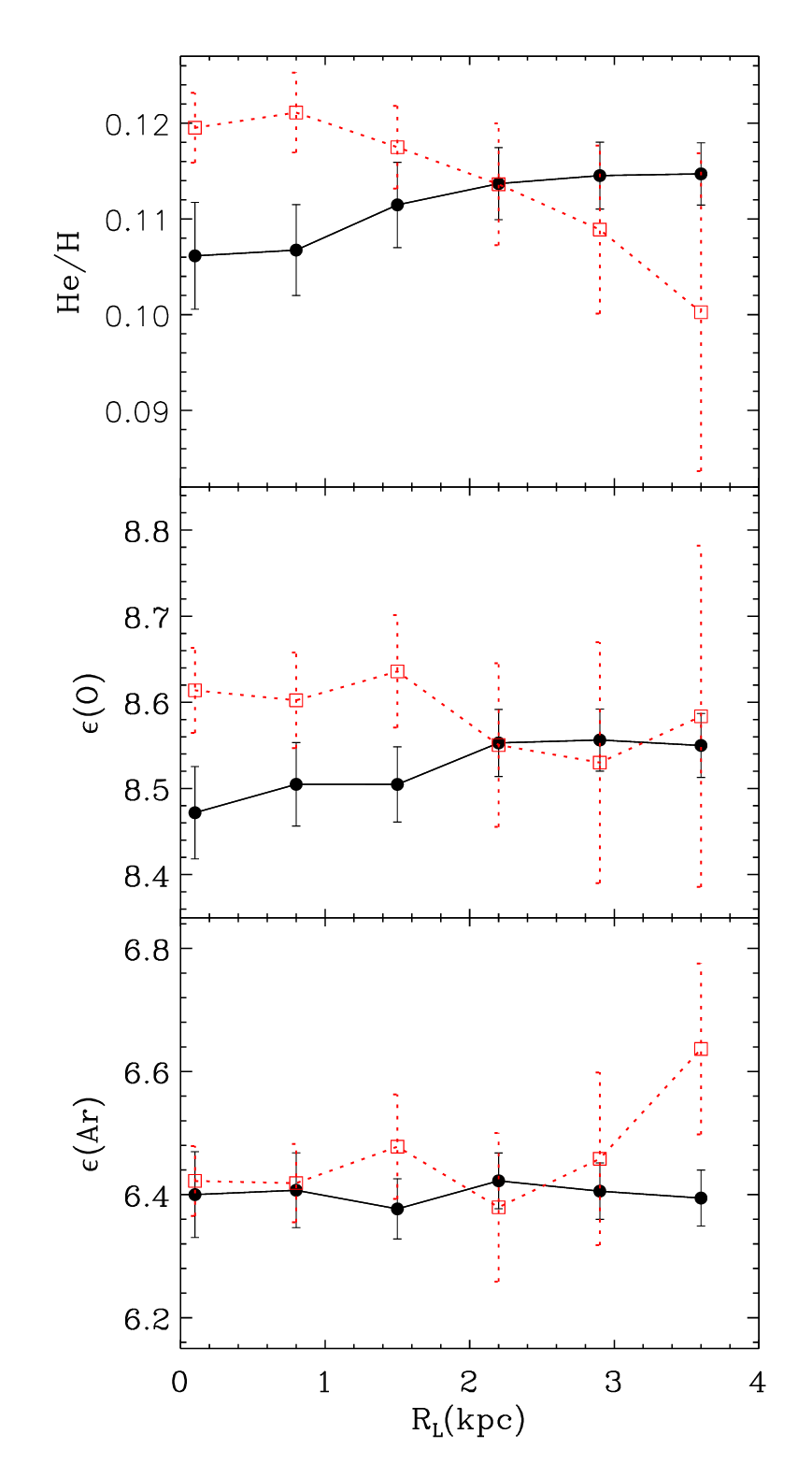}
	\includegraphics[width=6.5cm]{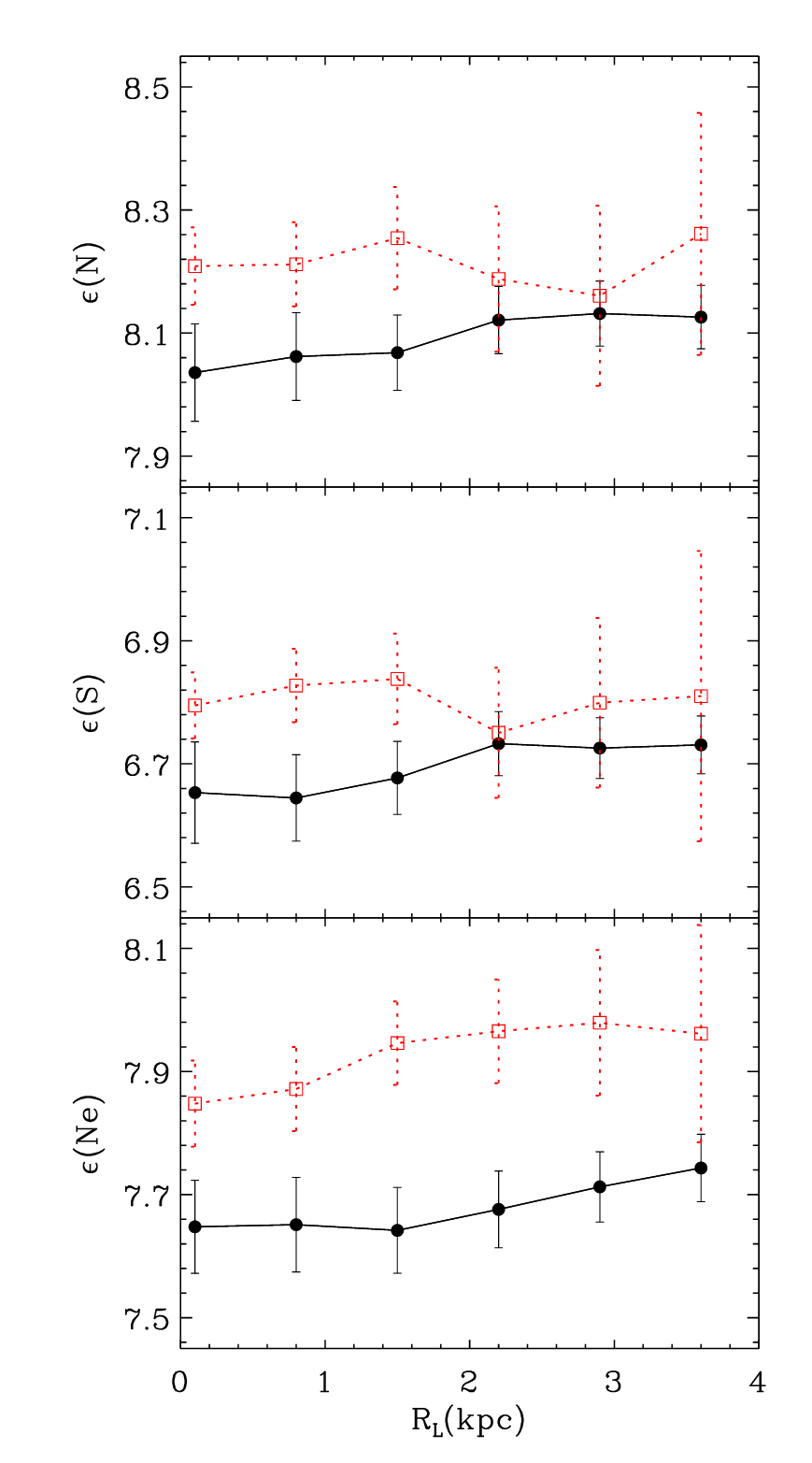}
	\caption{The same as figure \ref{stangh} for the Z95 scale.}
	\label{zhang}
\end{figure*}

The results are shown in figures \ref{stangh} and \ref{zhang} for SSV08 and Z95 distance scales, respectively. In each figure, group I is represented by filled circles joined by continuous lines and group II by squares joined by dashed lines. Each pair of circle/square points in every plot represents the average abundance for that element adopting a given limit for the bulge-disk interface, as a function of the limiting distance. Therefore, each plot shows how the differences between the groups evolve adopting distinct limits to define them. Considering elements heavier than He, it can be seen that, on average, abundances of group I are lower than group II objects for both distance scales. That is, in most cases the objects of group I, which are closer to the galactic center, have systematically lower abundances compared with the objects of group II, for all chosen values of the limiting radius  ($\mbox{R}_{\mbox{L}}$). The main exceptions are the helium abundances for both distance scales and argon for the Z95 scale, which present similar distributions for both groups. In the case of helium, these results are not surprising, as (A) this element is equally contaminated by the PN progenitor stars, both in the bulge and in the galactic disk, and (B) helium does not show any measurable radial abundance gradient, in contrast with the remaining elements considered in this paper. (See a detailed discussion of the helium gradient in \citet{maciel00}).

Examining figure \ref{stangh} we see that for the $\alpha$-elements oxygen, sulfur and argon the differences between both samples reach a minimum between 1.5  and 2.5 kpc, where it is of the order of 0.1 dex or lower. In figure \ref{zhang} examining now the $\alpha$-elements, we see that a minimum difference appears around 1.5 or 2 kpc. For neon, the abundance difference between the groups is of the order of 0.2 dex, while for the other elements this difference is of the order of 0.15 dex. For argon this difference is not clear, which is already expected since the ionization correction factor used to derive its abundance can lead to uncertainties larger than those for other elements.

\subsection{Kolmogorov-Smirnov test}

\begin{figure}[ht!]
	\centering
	\includegraphics[width=15cm]{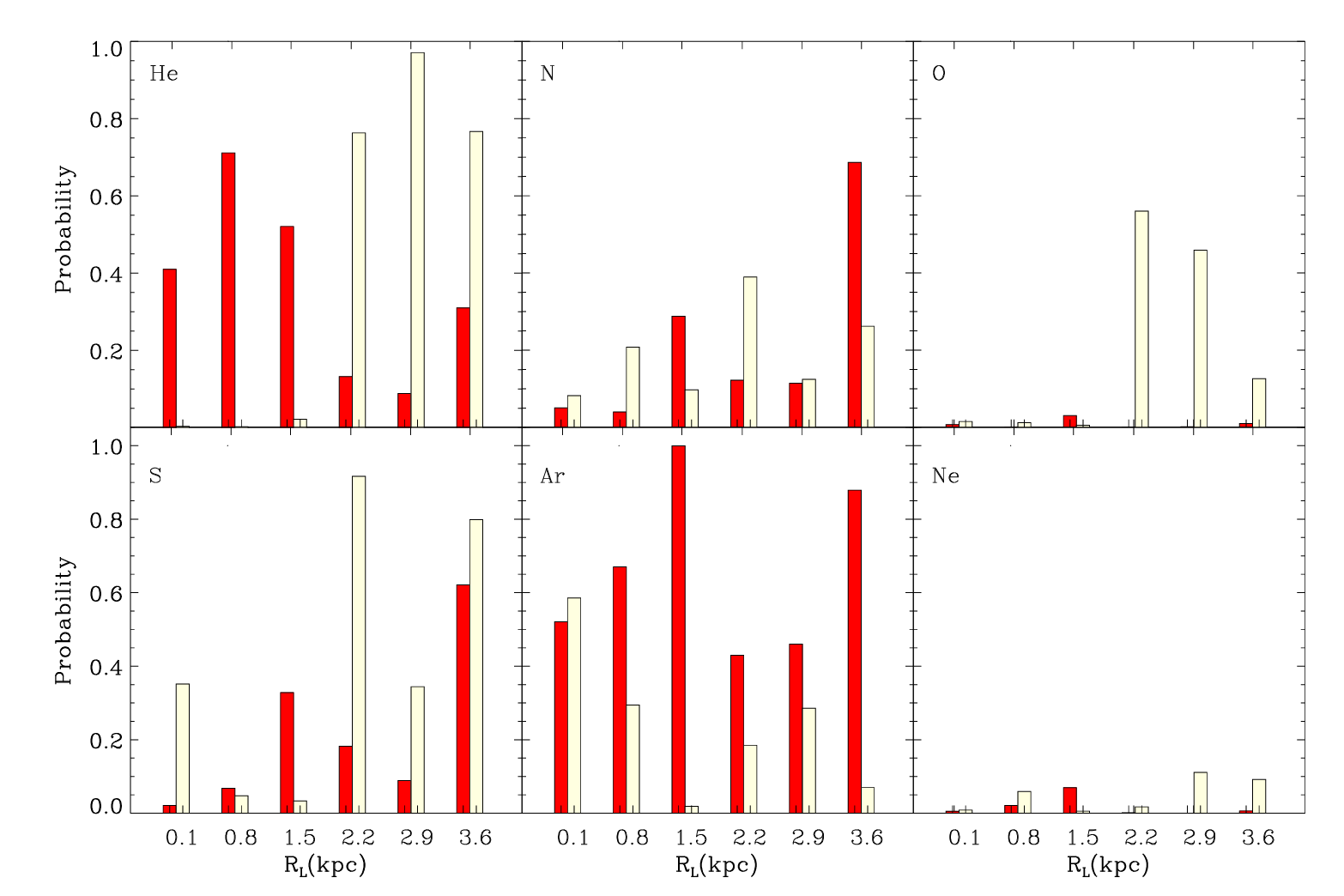}
	\caption{Kolmogorov-Smirnov probability bar plot for each element as a function of the limiting radius of groups I and II. Left filled bars correspond to the SSV08 scale and empty right bars to the Z95 scale.}
	\label{kolmo_stangh_zhang}
\end{figure}

Our goal in this section is to find the distance that better characterizes the separation of the two groups, in the sense that the populations of each group are the most distinct as possible. In this case, we will be able to ascertain a value for the galactocentric distance of the bulge-disk interface based on the intermediate mass population represented by the PNe sample.  In order to find such distance, we performed a Kolmogorov-Smirnov test for groups I and II at each step of the procedure described above. The test returns the probabilities that the two groups are drawn from the same distribution. Small probabilities show that the cumulative distribution function of group I is significantly different from group II. The results are displayed in figure \ref{kolmo_stangh_zhang}. For each element the probability is shown as function of the limiting distance $\mbox{R}_{\mbox{L}}$. For each distance step, the left filled bar represents the SSV08 scale and the right empty bar the Z95 scale. Since we are looking for small probabilities, from the figure we can see that the lowest probabilities are achieved at 1.5 kpc for the Z95 distance scale for the elements O, S, Ar, and Ne. The probabilities for each element and distance are listed in table \ref{tab_kolmo} and the lowest values are highlighted in black. Again, it is clear that for the Z95 distance scale the lower probabilities are achieved at 1.5 kpc for all the $\alpha$-elements. On the other hand, SSV08 does not show a unique distance for the lowest probabilities. Indeed, considering only the $\alpha$-elements, no distances are alike. Nevertheless, examining figure \ref{stangh} in detail, we can see that there is a step in the abundances radial distributions for the $\alpha$-elements concerning the distance of 2.2 kpc. So that the safest conclusion we can draw is that  $\mbox{R}_{\mbox{L}} < 2.9 \mbox{ kpc}$ based on the SSV08 distance scale.        

\begin{table}[ht!]
  \begin{center}
  \caption{Kolmogorov-Smirnov probabilities.}
  \label{tab_kolmo}
 {\scriptsize
  \begin{tabular}{@{\extracolsep{-0.05in}}l c c c c c c c c c c c c}
  \toprule
 	&  \multicolumn{6}{c}{ P$(\mbox{R}_{\mbox{L}})$ SSV08} 			&				\multicolumn{6}{c}{P$(\mbox{R}_{\mbox{L}})$ Z95} \\[-.20cm]\\
 \cmidrule(r){2-7}  \cmidrule(l){8-13}

$\mbox{R}_{\mbox{L}}$ (kpc)	&	0.1		&	0.8		&	1.5		&	2.2		&	2.9		&	3.6		&	0.1	&	0.8	&	1.5	&	2.2	&	2.9	&	3.6	\\[0.1cm]
																									
He						&	0.574	&	0.865	&	0.379	&	0.118	&	{\bf0.080}	&	0.293	&	0.003	&	{\bf 0.0026}	&	0.022	&	0.763	&	0.971	&	0.767	\\
																									
N						&	{\bf0.053}	&	0.075	&	0.335	&	0.103	&	0.099	&	0.647	&	{\bf 0.0819}	&	0.208	&	0.097	&	0.390	&	0.124	&	0.262	\\
																									
O						&	0.014	&	{\bf0.001}	&	0.058	&	{\bf0.001}	&	{\bf0.001}	&	0.008	&	0.015	&	0.012	&	{\bf 0.0051}	&	0.560	&	0.459	&	0.126	\\
																									
S						&	{\bf0.022}	&	0.074	&	0.370	&	0.154	&	0.077	&	0.583	&	0.352	&	0.047	&	{\bf 0.0334}	&	0.917	&	0.344	&	0.799	\\
																									
Ar						&	0.720	&	0.864	&	0.989	&	{\bf0.394}	&	0.508	&	0.845	&	0.586	&	0.295	&	{\bf 0.0188}	&	0.185	&	0.285	&	0.070	\\
																									
Ne						&	0.014	&	0.046	&	0.128	&	0.001	&	{\bf0.000}	&	0.007	&	0.009	&	0.059	&	{\bf 0.0056}	&	0.017	&	0.111	&	0.092	\\
\bottomrule
  \end{tabular}
  }
 \end{center}
\end{table}

\subsection{Resulting abundance distributions \label{sec:abund_distr}}

\begin{figure}[ht!]
	\centering
	\includegraphics[width=10cm]{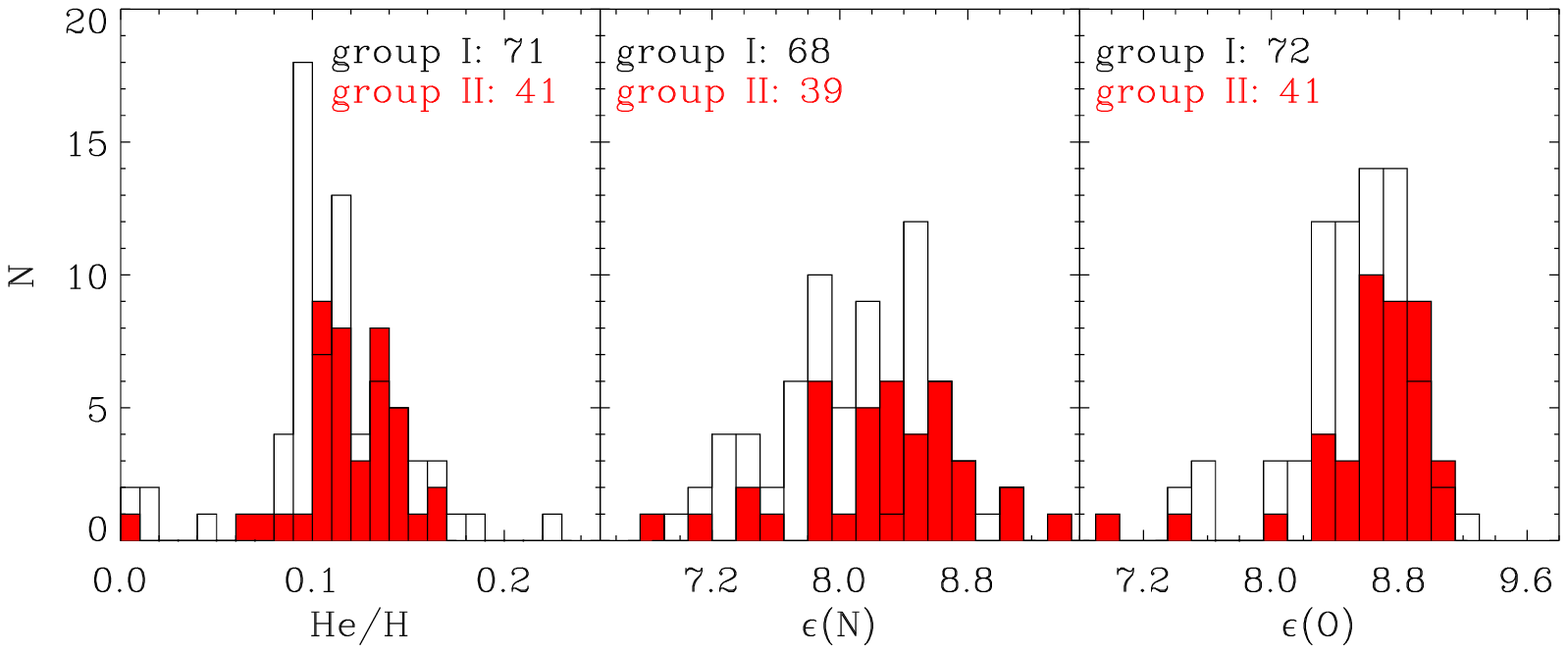}
	\includegraphics[width=10cm]{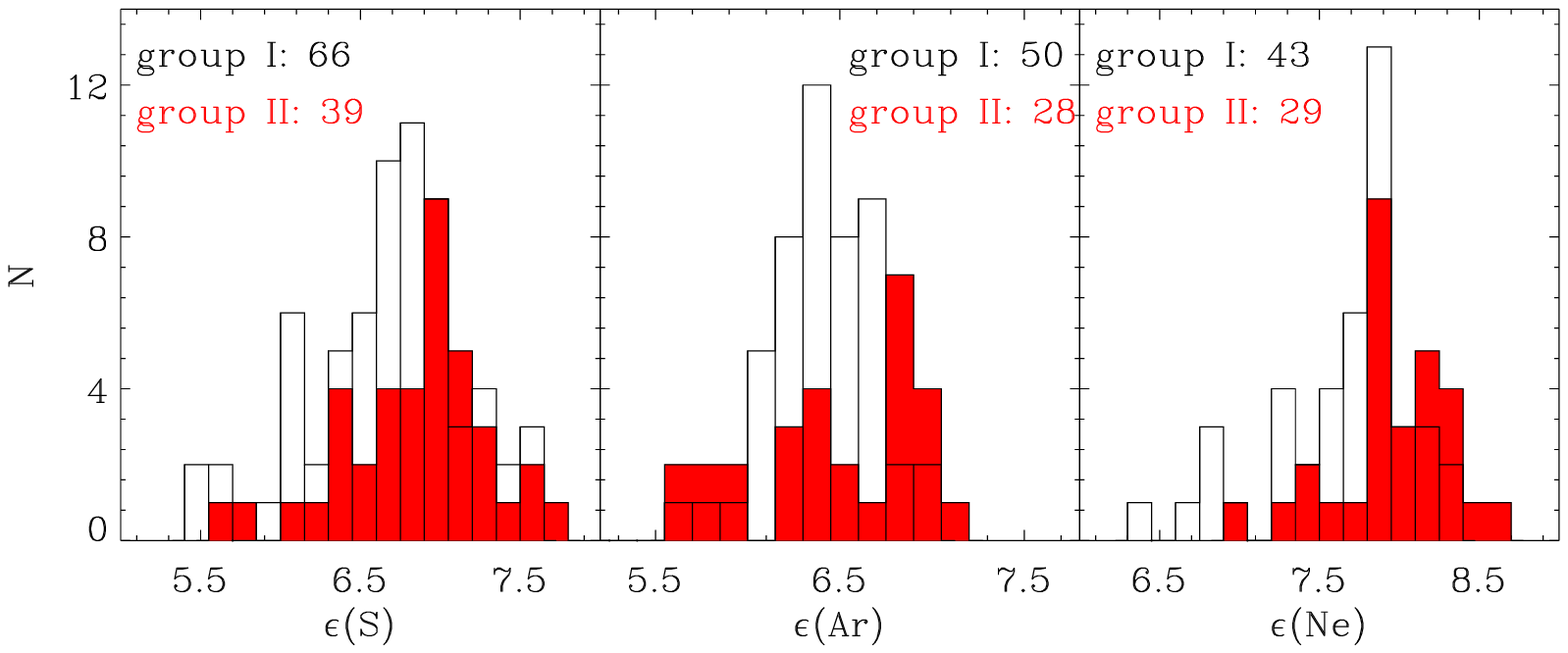}
	\caption{Abundance distributions for groups I and II using the Z95 distance scale. Unfilled histograms represent group I objects and filled histograms are for group II. The number of objects in each distribution is shown in the top.}
	\label{hist_zhang}
\end{figure}

\begin{figure}[ht!]
	\centering
	\includegraphics[width=10cm]{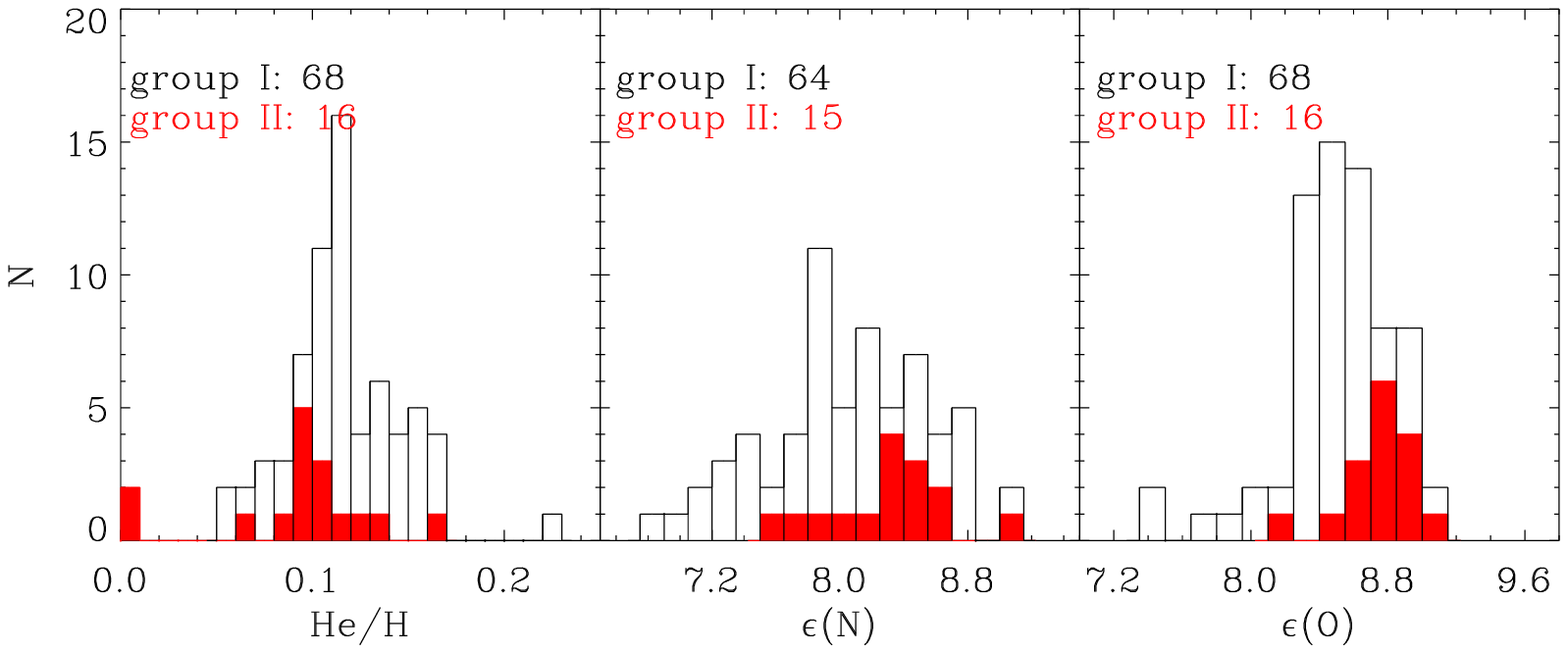}
	\includegraphics[width=10cm]{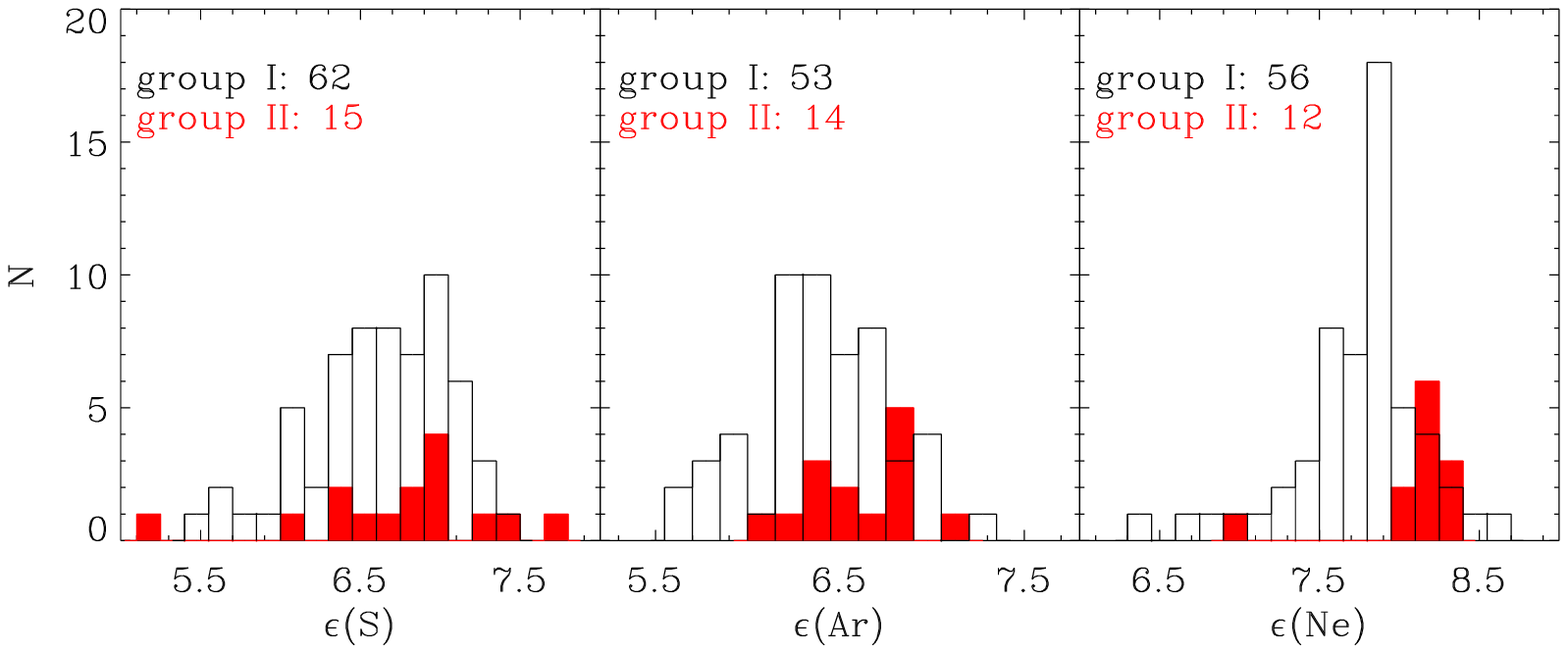}
	\caption{The same as figure \ref{hist_zhang} for SSV08 distance scale.}
	\label{hist_stangh}
\end{figure}

Adopting the separation distances obtained for each element and distance scale, namely 1.5 kpc for Z95 and 2.2 kpc for SSV08, the distribution of the abundances for each group are shown in figures \ref{hist_zhang} and \ref{hist_stangh} for Z95 and SSV08 distance scales, respectively. Empty histograms represent group I objects, while filled histograms represent group II objects. The number of objects used is also given for each plot.

In figure \ref{hist_zhang} the oxygen distribution for group I is wider than group II, while most objects from group II have abundances centered at 8.5 dex. The other $\alpha$-elements also show this pattern according to which group II objects favor higher abundances than those from group I, however these differences are not superior to the individual errors in abundances. Among the $\alpha$-elements, neon displays the largest difference between the two distributions. Nevertheless, some caution must be taken here. In a recent study by \citet{milingo10}, there are some evidences that neon enhancement is present in a significant portion of PNe. This enhancement could be explained considering the charge exchange reaction $\mbox{O}^{++}+\mbox{H}^0 \rightarrow \mbox{O}^+ + \mbox{H}^+$ \citep[see][and references therein]{peimbert95}. Since the neon ICF depends inversely on the  $\mbox{O}^{+ +}$ abundance, the charge exchange reaction could increase the neon elemental abundance. If the neon enhancement is real, it does not represent the abundance of the interstellar medium at the time of the formation of the progenitor star. Thus, the neon distribution in figure \ref{hist_zhang} must be interpreted carefully.

In figure \ref{hist_stangh}, that shows the distribution of the abundances for the SSV08 scale, the oxygen distribution for group I objects is wider than group II and the values are between 8.0 and 9.3 dex. The same trend is seen in the distributions for SSV08 scale when compared with the Z95 scale: compared to group I objects, the $\alpha$-element distributions for group II show a tendency to higher abundances. 

\begin{table}[ht!]
  \begin{center}
  \caption{Average abundances of groups I and II for each distance scale.}
  \label{tab_abund_groups}
 {\scriptsize
  \begin{tabular}{l c c c c }
  \toprule
        &  \multicolumn{2}{c}{SSV08} &    \multicolumn{2}{c}{Z95}  \\ 
    \cmidrule(r){2-3} \cmidrule(l){4-5} 
    Element             & Group I          &  Group II             & Group I &   Group II\\ 
   \midrule              
   He/H          & 0.110 $\pm$ 0.004 &  0.099 $\pm$ 0.009	& 0.111 $\pm$ 0.004 & 0.117 $\pm$ 0.004\\
   $\epsilon$(N) & 8.02 $\pm$ 0.06   &  8.22 $\pm$ 0.10  	&  8.07 $\pm$ 0.06  &	8.25 $\pm$ 0.08\\
   $\epsilon$(O) & 8.49 $\pm$ 0.04   &  8.69 $\pm$ 0.06  	&  8.50 $\pm$ 0.04  &	8.64 $\pm$ 0.07\\  	   
   $\epsilon$(S) & 6.62 $\pm$ 0.05   &  6.74 $\pm$ 0.14  	&  6.68 $\pm$ 0.06  &	6.84 $\pm$ 0.07\\
   $\epsilon$(Ar)& 6.37 $\pm$ 0.05   &  6.56 $\pm$ 0.07  	&  6.38 $\pm$ 0.05  &	6.48 $\pm$ 0.08\\
   $\epsilon$(Ne)& 7.74 $\pm$ 0.05   &  8.00 $\pm$ 0.12  	&  7.64 $\pm$ 0.07  &   7.95 $\pm$ 0.07\\ 
  \bottomrule
  \end{tabular}
  }
 \end{center}
\end{table}

Table \ref{tab_abund_groups} shows the average abundances and the standard error of the mean for group I and II objects, adopting the distances 1.5 and 2.2 kpc for the Z95 and SSV08 scales as the limit between the two groups. Although these abundances are similar, reflecting the large dispersion of abundances found in these regions of the Galaxy, some important differences are apparent. In particular, the N, O, S, Ar, and Ne abundances are {\it smaller} for group I compared to group II for both distance scales. For N and O the difference between the two groups is 0.20 dex. For S, Ar, and Ne it is 0.12, 0.19 and 0.26 dex, respectively. In spite of the fact that these differences are not larger than the individual errors in the abundances, we can draw important conclusions about these results. The standard error of the mean is small since both groups have several objects and therefore the difference between both groups is statistically significant . This can be seen considering the lower and upper 90\% and 68\% confidence limits of each group.  Although there are some overlaps between the distributions, for N, O, Ar, and Ne they are statistically significantly different considering the 68\% confidence limit of each group for the SSV08 distance scale. Just the S distributions are not statistically different considering this method. For the Z95 distance scale, the distributions of N, O, and S of each group are significantly different considering the same confidence limit. The Ne distributions are statistically different considering the 90\% confidence limit for this scale. These results are an important indication that bulge nebulae (group I) do not follow the trends observed in the inner-disk (group II), as we will discuss in section \ref{sec:discussion}.

Relaxing the assumption that objects with adopted negative galactocentric distances belong to the bulge would not change the derived conclusions. We performed some tests excluding these objects, and the main conclusions are unchanged: bulge objects have lower $\alpha$-element abundances than those in the inner disk. Adopting positive distances for these objects makes the
group separation less clear, as the mixing of both populations is enhanced, but even in this case
bulge objects still have lower abundances compared with inner disk objects.

The galactocentric distance of 1.5 kpc for Z95 distance scale mark the limit at which the groups I and II are more likely to separate from the point of view of the chemical abundances, as shown by the Kolmogorov-Smirnov test. On the other hand, the galactocentric distance of 2.2 kpc for the SSV08 scale was chosen from eye inspection in the radial abundances distributions, marking the step in the $\alpha$-elements abundances. In this case too, objects from group II, whose galactocentric distances are greater than 2.2 kpc, have higher abundances than those from group I, whose distances are lower than this limit. 

Since the bulge and the disk display different chemical abundance characteristics, as for example the radial abundance gradients found in the disk, or the large abundance distribution found in the bulge, as we discussed in section \ref{sec:intro}, we expect the existence of a galactocentric distance which better separates the two populations. Our analysis point to a separation at $\mbox{R}_{\mbox{L}} = 1.5 \mbox{ kpc}$. for Z95 distance scale and $\mbox{R}_{\mbox{L}} < 2.9 \mbox{ kpc}$ for the SSV08 scale. Therefore, we can assign to group I those objects that pertain to the bulge, and to group II those that pertain to the inner-disk.

\section{Discussion}
\label{sec:discussion}

There are other evidences that point to a transition from the bulge to the inner-disk at a galactocentric distance similar  to the distances found in this work. Indeed, infrared observations of the bulge, like those from the COBE/DIRBE satellite \citep{weiland94}, support these results. From their figure 1, it is clear that the nuclear radius of the bulge is between 10$^{\circ}$ and 15$^{\circ}$. For the canonical solar distance to the galactic center of 8 kpc adopted in this work, the bulge longer radius is between 1.4 and 2.1 kpc. 

From the point of view of giant stars, \citet{tiede99} did photometric and spectroscopic observations of 503 stars in four windows toward the bulge, with coordinates ($\ell$,b) of ($-28.8^\circ$,$-6^\circ$), ($-8.7^\circ$,$6.0^\circ$), ($8.4^\circ$,$-6.0^\circ$) and ($24.4^\circ$, $6.1^\circ$). According to their results, the abundance distribution in the inner Galaxy presents a discontinuity. The authors argue that such difference is larger than any bias in the sample selection or errors in the abundances. As we are suggesting, they assign the difference found to the different populations in the line of sight.  

\citet{smartt01} observed type-B stars located between 2.5 and 5 kpc from the galactic center. Due to the young nature of these stars, their photospheric abundances reflect the present day abundances of the ISM. Their results agree with ours in the sense that the oxygen abundance decreases for distances smaller than 3 kpc, as can be seen from their figure 3. However, the error bars are large and it is only possible to say that the oxygen abundances do not follow the radial gradient of the disk toward the galactic center, since they are 0.3 to 0.4 dex lower than what we expected from the \citet{rolleston00} radial gradient, given their position.

In the light of recent studies using PNe, \citet{guten08} point to a discontinuity in the abundance gradient towards the galactic center, in the sense that the abundances of bulge PNe do not follow the trend of those from the disk, as can be seen in their figure 4. The same trend can be seen in figure 3 of \citet{stangh10}. Considering PNe type III as defined by \citet{peimbert78}, which represent the old disk population, it is possible to note from their figure that oxygen abundances are smaller for galactocentric distances less than 4 kpc. 

Finally, \citet{mishurov02} proposed a galactic formation model where the spiral arms are inductors of stellar formation and they can exist only between the internal and external Lindblad resonances. As a consequence, the SFR decreases strongly outside these limits, and the chemical abundances are depleted in these regions. Combining these evidences with our results, we can point out to a bulge-disk interface for the intermediate mass population, marking therefore the transition between the disk and the bulge populations, at a galactocentric distance of 1.5 for the Z95 distance scale. On the other hand, for the SSV08 distance scale the Kolmogorov-Smirnov test was not conclusive, and we can point out that, for galactocentric distances lower than 2.9 kpc, the $\alpha$-elements abundances do not follow the disk radial gradient towards the center of the Milk Way galaxy.

\section{Summary and Conclusion}
\label{sec:summary}

This work reports an important result concerning PNe and the chemical evolution of the Galaxy. In addition to our previous work, \citep[][Paper I]{cavichia10}, where we presented the chemical abundances and the analysis of these abundances for a sample of 56 PNe in the direction of the galactic bulge, among which 35 PNe have their abundances derived for the first time, we extended our data base to 140 PNe in order to study the chemical abundance distribution in the inner disk and bulge of the Galaxy.  

In this work, a statistical analysis was performed in order to find the galactocentric distance where bulge and disk characteristics intersect. Two distance scales were used: \citet{zhang95} and \citet{stangh08}. Applying the Kolmogorov-Smirnov test, the former results in a distance of 1.5 kpc, while the latter the test was not conclusive. Nevertheless, the radial $\alpha$-elements abundance distributions for both scales indicate that they do not follow the trend of the disk. On average, the abundances for objects with galactocentric distances lower than 1.5 and 2.2 kpc for Z95 and SSV08 distance scales, respectively, are lower than objects with distances greater than this limit, although this difference is not larger than the errors in individual abundances. Considering the SSV08 distance scale, the abundance difference between the two groups of PNe is based on  68, 62, 53, and 56 PNe of group I (bulge) for O, S, Ar, and Ne, respectively,  and 16, 15, 14, 12 PNe of group II (inner disk) for the same elements. For the Z95 distance scale, the abundance difference is based on 72, 66, 50, and 43 PNe of group I, and 41, 39, 28, 29 for group II, for the same elements. Furthermore the combined error of the four elements (O, S, Ar, Ne) of each group is considerably smaller than the abundance difference between the two groups derived from the average difference of the four elements.

Taking into account the results derived in this work as well as other evidences from the literature, we propose the galactocentric distance of 1.5 kpc for the Z95 distance scale, to mark the transition between the bulge and inner-disk of the Galaxy.


\bigskip
{\it Acknowledgements.
      Part of this work was supported by the Brazilian agencies \emph{FAPESP} and \emph{CNPq}. O.C. would like to acknowledge FAPESP
for his graduate fellowship (processes 05/03194-4 and 07/07704-2). We also thank the referee for his/her comments and suggestions.
} 

\bibliography{biblio}

\end{document}